\documentclass[11pt]{article}
\usepackage{hyperref}
\usepackage{amsmath}
\usepackage{amssymb,amsthm}
\usepackage[english]{babel}
\usepackage[textwidth=18cm,textheight=22cm]{geometry}
\usepackage{graphicx}
\usepackage{psfrag}
\usepackage{array}

\renewcommand{\d}{\mathrm{d}}
\newcommand{\bt}{\boldsymbol{t}}

\newcommand{\bF}{\mathbb{F}}
\newcommand{\bV}{\mathbb{V}}

\newtheorem{teh}{Theorem}

\newcommand{\be}{\begin{equation}}
\newcommand{\ee}{\end{equation}}
\begin{document}

\title{\sc String equations in  Whitham hierarchies:\\  $\tau$-functions and Virasoro constraints\thanks{Partially supported by DGCYT
project BFM2002-01607 }}
\author{Luis Mart\'{\i}nez Alonso$^{1}$,
 Elena Medina$^{2}$ and Manuel Ma\~{n}as$^{1}$\\\\
\emph{ $^1$Departamento de F\'{\i}sica Te\'{o}rica II, Universidad
Complutense}\\ \emph{E28040 Madrid, Spain} \\
\emph{$^2$Departamento de Matem\'{a}ticas, Universidad de
C\'{a}diz}\\\emph{ E11510, Puerto Real, C\'{a}diz, Spain}}

\date{}
\maketitle

\begin{abstract}
A  scheme for solving Whitham hierarchies satisfying a special class of string equations is
presented. The $\tau$-function of the corresponding solutions is obtained
and the differential expressions of the underlying Virasoro constraints are characterized.
Illustrative examples of exact solutions of Whitham hierarchies are derived and applications
to conformal maps dynamics are indicated.

\end{abstract}

\vspace*{.5cm}

\begin{center}\begin{minipage}{12cm}

\end{minipage}
\end{center}
\newpage

\section{Introduction}

Nonlinear integrable models of dispersionless type
\cite{h1}-\cite{krich1} arise  in several branches of physics and applied mathematics.
They have gained prominence
after the discovery of their relevance in  the formalism of quantum
topological fields \cite{krich2}-\cite{kod}, and of their role in the
theory of deformations of conformal and
quasiconformal maps on the complex plane \cite{gib}-\cite{kon2} .
Recently,  new applications  have been found
\cite{zab1}-\cite{zab7} which include  dynamics of conformal maps,
growth processes of Laplacian type and large $N$ limits of
random matrix partition functions.

 From the point of view of the theory of integrable systems , these
 models turn to be furnished by members of the so called
\emph{universal Whitham hierarchies} introduced by Krichever in
\cite{krich1}-\cite{krich2}. A particularly  important example of these hierarchies
is the dispersionless Toda (dToda) hierarchy \cite{zab1}-\cite{zab4}, \cite{kaz}-\cite{tak1}.
 The  solutions of dispersionless integrable models
 underlying  many of their applications satisfy special systems of
constraints called \emph{string equations},  which posses
attractive mathematical properties and interesting physical
meaning.  Takasaki and Takebe \cite{tt1}-\cite{tt4} showed the relevance of  string equations for studying
the dispersionless KP and Toda hierarchies  and, in particular, for characterizing their
associated symmetry groups. Nevertheless,
 although  some schemes for
solving  string equations in the dispersionless KP and Toda hierarchies were
provided in \cite{mel1}-\cite{mel3}, general efficient methods of solution for string equations are still lacking.

In a recent work \cite{mano} a general formalism
of Whitham hierarchies based on a factorization problem
on a Lie group of canonical transformations has been proposed. It leads to a natural
formulation of string equations
in terms of dressing transformations.  The  present  paper is concerned with the analysis of
these string equations and, in particular, their applications for characterizing exact solutions
of Whitham hierarchies. Thus we provide a solution scheme for
an special class of  string equations which  determines not only the solutions of
the algebraic orbits of
the Whitham hierarchy \cite{krich2} , but also the solutions
arising in the above mentioned applications of dispersionless
integrable models \cite{mel2}-\cite{mel3}. We characterize the
$\tau$-function  corresponding to these solutions and, by taking advantage of the string
equations, we also derive  the differential expressions of the underlying Virasoro constraints.

 The elements of the phase space for a zero
genus Whitham hierarchy are characterized by a finite set
\[
(q_{\alpha},z_{\alpha}^{-1}(p)),\quad \alpha=0,\ldots,M,
\]
of \emph{punctures}  $q_{\alpha}$, where $q_0:=\infty$, of the
complex $p$-plane  and an associated set of local coordinates of
the form
\begin{gather}\label{1.1}\everymath{\displaystyle}
      z_\alpha=\begin{cases}
        p+\sum_{n=1}^\infty \frac{d_{0n}}{p^n}, & \alpha=0,\\\\
        \dfrac{d_{i}}{p-q_i}+\sum_{n=0}^\infty d_{in} (p-q_i)^n,&
        \alpha=i=1,\dots, M.
      \end{cases}
\end{gather}
The set of flows of the Whitham hierarchy can be formulated as the
following infinite system of quasiclassical Lax equations
\begin{equation}\label{wh} \frac{\partial z_{\alpha}}{\partial
t_{\mu n}}=\{\Omega_{\mu n}, z_{\alpha}\},
\end{equation}
where the Poisson bracket is defined as
\[
\{F,G\}:=\frac{\partial F}{\partial p} \frac{\partial G}{\partial
x} -\frac{\partial F}{\partial x}\frac{\partial G}{\partial p}.
\]
and the Hamiltonian functions are
\begin{gather}\label{1.3}\everymath{\displaystyle}
\Omega_{\mu n}:=\begin{cases}   (z_\mu^n)_{(\mu,+)} ,& n\geq 1
,\\\\
-\log_i(p-q_i), &n=0,\quad \mu=i=1,\dots,M.
\end{cases}
\end{gather}
Here $(\cdot)_{(i,+)}$ and $(\cdot)_{(0,+)}$ stand for the
projectors on the subspaces generated by
$\{(p-q_i)^{-n}\}_{n=1}^\infty$ and $\{p^n\}_{n=0}^\infty$ in the
corresponding spaces of Laurent series. Henceforth, it will be
assumed that appropriate non-intersecting cuts connecting
$p=\infty$ with the points $q_i$ are made which allow us to define
the logarithmic branches associated with $\Omega_{i0}$. Since
several of these branches will appear simultaneously in certain
equations, to avoid possible misunderstanding we introduce the
notation convention $\log_i(p-q_i)$. For $M=0$ and $M=1$ these
systems represent the dispersionless versions of the KP and Toda
hierarchies, respectively.

In what follows Greek and Latin suffixes will be used to label
indices of the sets $\{0,\ldots,M\}$ and $\{1,\ldots,M\}$,
respectively. In our analysis we use an extended Lax formalism
with  Orlov functions
\begin{equation}\label{1.2}
m_{\alpha}(z,\bt)=\sum_{n=1}^{\infty}nt_{\alpha
n}z_{\alpha}^{n-1}+\frac{t_{\alpha 0}}{z_{\alpha}}+
\sum_{n\geq2}\frac{v_{\alpha n}}{z_{\alpha}^n},\quad t_{0
0}:=-\sum_{i=1}^Mt_{i 0},
\end{equation}
such that
\[
\{z_\alpha,m_\alpha\}=1,\quad \forall \alpha,
\]
and verifying the same Lax equations \eqref{wh} as the variables
$z_\alpha$.

The basic notions about the Whitham hierarchy which are necessary
for the subsequent discussion  are introduced  in Sec. 2.  String
equations and symmetries are discussed in Sec.3, where
the main results concerning the construction of solutions from
meromorphic string equations and their Virasoro invariance  are proved. Section 4 presents a scheme for
solving an special class of  string equations,  which is
illustrated with several explicit  examples. A
formula for the corresponding $\tau$-function is given which generalizes the expression of the $\tau$-function of
analytic curves found in \cite{zab1}. Finally, we analize the Virasoro symmetries associated to the string equations and  obtain
the corresponding Virasoro constraints in differential form.

\section{The Whitham hierarchy}

In order to display the main features of the Whitham hierarchy it is convenient to use the following concise formulation in terms of the  system of equations
\begin{equation}\label{2.a}
\d z_{\alpha}\wedge\d m_{\alpha}=\d \omega,\quad \forall \alpha,
\end{equation}
where $\omega$ is the one-form defined by
\begin{equation}\label{2.aaa}
\omega:=\sum_{\mu,n}\Omega_{\mu n} \d t_{\mu n}.
\end{equation}
To see how to get from the system \eqref{2.a} to the  Whitham hierarchy, note that by identifying the coefficients of $\d p\wedge\d t_{\mu
n}$ and $\d x\wedge\d t_{\mu n}$ in \eqref{2.a} we obtain

\begin{equation}\label{2.b}\everymath{\displaystyle}
\begin{cases}
\frac{\partial z_{\alpha}}{\partial p}\frac{\partial
m_{\alpha}}{\partial t_{\mu n}} -\frac{\partial
m_{\alpha}}{\partial p}\frac{\partial z_{\alpha}}{\partial t_{\mu
n}}&=\frac{\partial \Omega_{\mu n}}{\partial p},\\\\
\frac{\partial z_{\alpha}}{\partial x}\frac{\partial
m_{\alpha}}{\partial t_{\mu n}} -\frac{\partial
m_{\alpha}}{\partial x}\frac{\partial z_{\alpha}}{\partial t_{\mu
n}}&=\frac{\partial \Omega_{\mu n}}{\partial x}.
\end{cases}
\end{equation}
and, in particular, since $\Omega_{01}=p$, for $(\mu,n)=(0,1)$ ,
the system \eqref{2.b} implies
\[
\{z_{\alpha},m_{\alpha}\}=1.
\]
Thus, using this fact and solving  \eqref{2.b} for
$\dfrac{\partial z_{\alpha}}{\partial t_{\mu n}}$ and
$\dfrac{\partial m_{\alpha}}{\partial t_{\mu n}}$, we deduce
\[
\frac{\partial z_{\alpha}}{\partial t_{\mu n}}=\{\Omega_{\mu n},
z_{\alpha}\},\quad \frac{\partial m_{\alpha}}{\partial t_{\mu
n}}=\{\Omega_{\mu n}, m_{\alpha}\}.
\]

It is now natural to introduce the $S$-functions of the Whitham hierarchy.  Indeed as a consequence of \eqref{2.a} we find
\[
\d \Big(m_{\alpha}\,\d z_{\alpha} +\sum_{\mu,n}\Omega_{\mu n} \d
t_{\mu n}\Big)=0,\quad \forall \alpha,
\]
so that there exist functions $S_{\alpha}(z_{\alpha},\bt)$ such
that
\begin{equation}\label{2aaa}
\d S_{\alpha}=m_{\alpha}\,\d z_{\alpha}
+\sum_{\mu,n}\Omega_{\mu n} \d t_{\mu n},\quad \forall \alpha,
\end{equation}
and from \eqref{1.2} we see that they admit expansions of the form
\begin{equation}\label{2aa}
S_{\alpha}=\sum_{n\geq 1}z_{\alpha}^{n} t_{\alpha n}+\log
z_{\alpha}t_{\alpha 0}-v_{\alpha} (\bt)-\sum_{n\geq
1}\frac{v_{\alpha n+1}}{n}\frac{1}{z_{\alpha}^{n}},\quad
z_{\alpha}\rightarrow\infty.
\end{equation}
It is important to notice that from \eqref{1.1}-\eqref{1.2} and
\eqref{2aaa} it follows that
\[
\d S_{0}=\sum_{n\geq 1}\Big(nz_{0}^{n-1}t_{0 n}\d z_{0}+(z_{0}^n)_{(0,+)}\d t_{0 n}\Big)+\frac{t_{0 0}}{z_{0}}\d z_{0}+\mathcal{O}\Big(\frac{1}{z_{0}^2}\Big)\d z_{0},\quad z_{0}\rightarrow\infty,
\]
and consequently we may take
\[
v_0(\bt)\equiv 0.
\]

\begin{figure}\label{fig 1}
\begin{center}
\psfrag{p}{$p$-plane}\psfrag{z}{$z_\mu$-plane}
\psfrag{a}{$\gamma_0$}\psfrag{b}{$\gamma_1$}\psfrag{c}{$\gamma_2$}
\psfrag{d}{$\gamma_3$} \psfrag{e}{$\Gamma_{\mu}$}
\includegraphics[width=13cm]{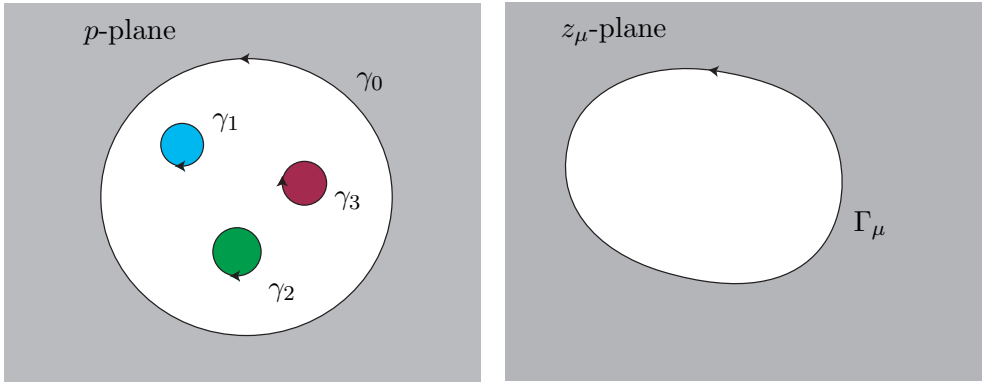}
\end{center}\caption{Right-exteriors of $\gamma_{\mu}$ and $\Gamma_{\mu}$}
\end{figure}

To proceed further  some analytic properties of the dynamical variables of the Whitham hierarchy are required. Thus we will henceforth suppose that there exist positively oriented closed
curves
$\Gamma_{\mu}$ in the complex planes of the variables $z_{\mu}$ such that each function $z_{\mu}(p)$
determines a conformal map of the right-exterior of a
circle $\gamma_{\mu}:=z_{\mu}^{-1}(\Gamma_{\mu})$  on the
exterior of $\Gamma_{\mu}$ (we will assume that the circle $\gamma_0$   encircles all the $\gamma_i,(i=1,\ldots M)$)  (see figure 1).
 Moreover, for each $\alpha$ the functions $S_{\alpha}$ and $m_{\alpha}$ will be assumed to be analytic in the exterior of $\Gamma_{\alpha}$.

 Under the above conditions
one can prove that
\begin{equation}\label{free 0}
\partial_{\beta,m}v_{\alpha,n+1}=\partial_{\alpha,n}v_{\beta,m+1},
\quad \forall \alpha, \beta;\quad  n, m\geq 0,
\end{equation}
where the functions $v_{i1}$ are defined by
\begin{equation}\label{new}
v_{i1}:=v_i-\sum_{j<i}\log_{ji}(-1)t_{j0},
\end{equation}
and we are denoting
\[
 \log_{ji}(-1):=\log_j(q_i-q_j)-\log_i(q_j-q_i)=-\log_{ij}(-1).
\]
 In other words, it is ensured the existence of a
\emph{free-energy function} $F=F(\bt)$ , the logarithm $F=\log\tau$
of the \emph{dispersionless} $\tau$-function, verifying
\begin{equation}\label{free}
\d F=\sum_{(\alpha,n)\neq (0,0)} v_{\alpha n+1}\d t_{\alpha n}.
\end{equation}

Let us first prove \eqref{free 0} for the case
$(\alpha,n)=(i,0),\;(\beta,m)=(j,0)$. From the equations
\[
\partial_{l0}S_k=-\log_l(p-q_l),
\]
it follows that
\[
\partial_{i0}v_j=\log_i(q_j-q_i),\quad
\partial_{j0}v_i=\log_j(q_i-q_j),
\]
so that the functions defined in \eqref{new} satisfy
\[
\partial_{i0}v_{j1}=\partial_{j0}v_{i1}.
\]
We indicate the strategy for proving the remaining cases of
\eqref{free 0} by considering the choice $\alpha=i,\;\beta=j\geq
1,\;n,m\geq 1$ of \eqref{free 0}. From \eqref{1.2} and \eqref{2aaa}
it follows that
\[
v_{i,n+1}=\frac{1}{2\pi i}\oint_{\Gamma_i}z_i^n m_i \d z_i,\quad
\partial_{j,m} S_i=(z_j^m)_{(j,+)},
\]
so that
\begin{align}\label{cross}
\nonumber \partial_{j,m}v_{i,n+1}&=\frac{1}{2\pi i}\oint_{\gamma_i}z_i^n\d (z_j^m)_{(j,+)}
=\frac{1}{2\pi i}\oint_{\gamma_i}(z_i^n)_{(i,+)}\d (z_j^m)_{(j,+)}\\\\
\nonumber &=\frac{1}{2\pi i}\oint_{\gamma_j}(z_j^m)_{(j,+)}\d
(z_i^n)_{(i,+)}=\partial_{i,n}v_{j,m+1},
\end{align}
where we have taken into account that $(z_i^n)_{(i,+)}\partial_p (z_j^m)_{(j,+)}$ is a rational
function of $p$ which has finite poles at $q_i$ and $q_j$ only and a zero
residue at $\infty$.

\section{String equations and symmetries}

As it was shown in \cite{mano},  the analysis of the factorization problem for the  Whitham
hierarchy  shows that this hierarchy admits a natural formulation in terms
of systems of string equations of the form
\begin{equation}\label{2.1}
\begin{cases}
P_i(z_i,m_i)=P_0(z_0,m_0),\\
Q_i(z_i,m_i)=Q_0(z_0,m_0),
\end{cases}\quad i=1,2,\dots,M,
\end{equation}
where $\{P_\alpha,Q_\alpha\}_{\alpha=0}^M$ are  pairs of canonically conjugate variables
\begin{equation}\label{2.2}
\{P_{\alpha}(p,x),Q_{\alpha}(p,x)\}=1,\quad \forall \alpha.
\end{equation}
In what follows we consider the problem of finding systems of the form \eqref{2.1} which
are appropriate to generate exact solutions of the Whitham hierarchy.

Given a solution $(z_{\alpha}(p,\bt),m_{\alpha}(p,\bt))$ of a system \eqref{2.1}, if we denote
\[
\mathcal{P}_{\alpha}(p,\bt):=
P_{\alpha}(z_{\alpha}(p,\bt),m_{\alpha}(p,\bt)),\quad
\mathcal{Q}_{\alpha}(p,\bt):=Q_{\alpha}(z_{\alpha}(p,\bt),m_{\alpha}(p,\bt)),
\]
then \eqref{2.1} and \eqref{2.2} imply
\begin{equation}\label{2.6}
\d \mathcal{P}_{\alpha}\wedge\d \mathcal{Q}_{\alpha}=\d \mathcal{P}_{\beta}\wedge\d \mathcal{Q}_{\beta},\quad \forall \alpha,\beta
\end{equation}
and
\begin{equation}\label{2.7}
\d \mathcal{P}_{\alpha}\wedge\d \mathcal{Q}_{\alpha}=\d z_{\alpha}\wedge\d
m_{\alpha},\quad \forall \alpha,
\end{equation}
respectively. Hence solutions of the system of string equations verify
\begin{equation}\label{2.8}
\d\mathcal{P}_{\alpha}\wedge\d\mathcal{Q}_{\alpha}=\d z_{\beta}\wedge\d m_{\beta},\quad \forall \alpha,\beta.
\end{equation}
The next result provides a convenient framework for our subsequent
discussion of  solutions of \eqref{2.1}.

\begin{teh}
Let $(z_{\alpha}(p,\bt),m_{\alpha}(p,\bt))$ be a  solution of \eqref{2.1} which admits  expansions
of the form  \eqref{1.1}-\eqref{1.2} and such that the coefficients of the
the two-forms \eqref{2.8}  are meromorphic functions of the complex variable $p$ with finite poles at
$\{q_1,\dots,q_M\}$ only. Then $(z_{\alpha}(p,\bt),m_{\alpha}(p,\bt))$
is a solution of the Whitham hierarchy.
\end{teh}

\begin{proof}

In view of the hypothesis of the theorem the coefficients of
the two-forms \eqref{2.8} with respect to the basis
\[
\{\d p\wedge\d t_{\alpha n},\quad\d t_{\alpha n}\wedge\d t_{\beta m} \}
\]
are determined by their  principal parts at $q_{\mu},
(\mu=0,\ldots ,M)$, so that by taking \eqref{2.8} into account we may write
\[
\d z_{\alpha}\wedge\d m_{\alpha}=\sum_{\mu=0}^M (\d z_{\mu}\wedge\d m_{\mu})_{(\mu,+)},\quad \forall \alpha.
\]
Moreover the terms in these decompositions  can be found by using the expansions \eqref{1.2} of the functions $m_{\mu}$ as follows
\begin{align*}
& \d z_{\mu}\wedge \d m_{\mu}=\d z_{\mu}\wedge\Big(
\sum_{n=1}^{\infty}nz_{\mu}^{n-1}\d t_{{\mu}n}+\frac{\d t_{{\mu}
0}}{z_\mu}+ \sum_{n\geq2}\frac{\d
v_{{\mu} n}}{z_{\mu}^n}\Big)\\
&=\d \Big(\sum_{n=1}^{\infty}z_{\mu}^{n}\d t_{{\mu}n}+\log z_{\mu}
\d t_{{\mu} 0}- \sum_{n\geq2}\frac{1}{n-1}\frac{\d
v_{{\mu}n}}{z_{\mu}^{n-1}}\Big),
\end{align*}
so that
\begin{align*}
(\d z_{\mu}\wedge\d m_{\mu})_{(\mu,+)}&=\d
\Big(\sum_{n=1}^{\infty}(z_{\mu}^{n})_{(\mu,+)}\d
t_{{\mu}n}-(1-\delta_{\mu 0})\log (p-q_{\mu}) \d
t_{{\mu}0}\Big)\\
&=\d\Big(\sum_{n}\Omega_{\mu n} \d t_{\mu n}\Big).
\end{align*}
Thus we find
\[
\d z_{\alpha}\wedge\d m_{\alpha}=\d\omega=\d \Big(\sum_{\mu,n}\Omega_{\mu n} \d t_{\mu n}\Big),\quad \forall \alpha,
\]
and, consequently, this proves that the functions  $(z_{\alpha}(p,\bt),m_{\alpha}(p,\bt))$
determine a solution of the Whitham hierarchy.
\end{proof}

Following the dressing scheme  of \cite{tt1}-\cite{tt4} it can be shown \cite{mano}
that each solution of the Whitham hierarchy is determined by an associated system of string equations.

As it was shown in \cite{mano} a complete formulation of the symmetry group of the Whitham hierarchy
is obtained by considering deformations of the associated factorization problem. On the other hand,
a natural representation of this group is provided
by the following symmetries of
string equations implemented by Hamiltonian vector fields:

\begin{teh}
Given  a vector function
\begin{equation}\label{sim1}
\bF:=(F_0(z_0,m_0),\ldots,F_M(z_M,m_M)),
\end{equation}
the infinitesimal deformation
\begin{align}\label{sim02}
\nonumber &\delta_{\bF}P_{\alpha}:=\{F_{\alpha},P_{\alpha}\},\quad
\delta_{\bF}Q_{\alpha}:=\{F_{\alpha},Q_{\alpha}\},\\\\
\nonumber &\delta_{\bF} z_{\alpha}:=\{ z_{\alpha},(F_{\alpha})_- \}
,\quad \delta_{\bF} m_{\alpha}:=\{ m_{\alpha},(F_{\alpha})_- \},
\end{align}
where
\[
(F_{\alpha})_-:=F_{\alpha}-\sum_{\beta}(F_{\beta})_{(\beta,+)},
\]
determines a symmetry of the system of string equations \eqref{2.1}.
\end{teh}
\begin{proof}
We have to prove that given a solution $(z_{\alpha},m_{\alpha})$ of \eqref{2.1}, then at first order in $\epsilon$
\begin{equation}\label{sim1a}
\begin{cases}
(P_i+\epsilon\;\delta_{\bF}P_i)(z_i+\epsilon\;\delta_{\bF}z_i,m_i+\epsilon\;\delta_{\bF}m_i)
=(P_0+\epsilon\;\delta_{\bF}P_0)(z_0+\epsilon\;\delta_{\bF}z_0,m_0+\epsilon\;\delta_{\bF}m_0),\\
(Q_i+\epsilon\;\delta_{\bF}Q_i)(z_i+\epsilon\;\delta_{\bF}z_i,m_i+\epsilon\;\delta_{\bF}m_i)
=(Q_0+
\delta_{\bF}Q_0)(z_0+\epsilon\;
\delta_{\bF}z_0,m_0+\epsilon\;\delta_{\bF}m_0),
\end{cases}
\end{equation}
for all $i=1,\ldots,M$. Let us consider the first group of equations of \eqref{sim1a}, they can be rewritten as
\[
\frac{\partial P_i}{\partial z_i}\delta_{\bF}z_i+\frac{\partial P_i}{\partial m_i}\delta_{\bF}m_i+\{F_i,P_i\}=
\frac{\partial P_0}{\partial z_0}\delta_{\bF}z_0+\frac{\partial P_0}{\partial m_0}\delta_{\bF}m_0+\{F_0,P_0\},
\]
or, equivalently, by taking \eqref{sim02} into account, as
\begin{equation}\label{sim03}
\{F_i-(F_i)_-,P_i\}=\{F_0-(F_0)_-,P_0\},\quad \forall i.
\end{equation}
By hypothesis $P_i(z_i,m_i)=P_0(z_0,m_0)$. On the other hand
\[
F_i-(F_i)_-=F_0-(F_0)_-=\sum_{\beta}(F_{\beta})_{(\beta,+)},
\]
so that \eqref{sim03} is satisfied. The proof for  the second group of equations of \eqref{sim1a} is identical.
\end{proof}

We note that the condition for a solution $(z_{\alpha},m_{\alpha})$ of the string equations \eqref{2.1} to be invariant under a symmetry \eqref{sim02} is
\begin{equation}\label{sim3}
(F_{\alpha}(z_{\alpha},m_{\alpha}))_-=0,\quad \forall \alpha,
\end{equation}
or equivalently
\begin{equation}\label{sim4}
F_{\alpha}=\sum_{\mu}(F_{\mu})_{(\mu,+)},\quad \forall \alpha.
\end{equation}
In other words,  the functions $F_{\alpha}(z_{\alpha},m_{\alpha})$ must reduce to a unique
meromorphic function of $p$ with
finite poles at the punctures $q_i$ only. As a consequence it follows that, under the hypothesis of Theorem 1, solutions of the Whitham hierarchy
satisfying a system of string equations \eqref{2.1} are invariant under the symmetries generated by
\[
\mathbb{P}=(P_0(z_0,m_0),\ldots,P_M(z_M,m_M)),\quad
\mathbb{Q}=(Q_0(z_0,m_0),\ldots,Q_M(z_M,m_M)),
\]
and, more generally, they are  invariant under the symmetries
generated by
\begin{equation}\label{sim5}
\bV_{rs}=(P_0^{r+1}\,Q_0^{s+1},\ldots,P_M^{r+1}\, Q_M^{s+1}),\quad
r\geq -1,\,s\geq 0,
\end{equation}
which determine a Poisson
Lie algebra $\mathcal{W}$ of symmetries
\[
\{\bV_{rs},\bV_{r's'}\}=((r+1)(s'+1)-(r'+1)(s+1))\bV_{r+r'\,
s+s'}.
\]
In particular the functions $\bV_r:=\bV_{r0}$ and $\tilde{\bV}_s:=-\bV_{0s}$ generate two Virasoro
algebras.
\[
\{\bV_{r},\bV_{r'}\}=(r-r')\bV_{r+r'},\quad
\{\tilde{\bV}_{s},\tilde{\bV}_{s'}\}=(s-s')\tilde{\bV}_{s+s'}.
\]

\section{A solvable class of string equations}

  In \cite{mano} a class of string equations was introduced which manifests special properties
with respect to the group of dressing transformations. We next provide a scheme of solution
for this class.

Let us consider systems of string equations associated to  splittings
\[
\{1,\ldots,M\}=I\cup J,\;
I\cap J=\emptyset
\]
of the form
\begin{equation}\label{3.1}\everymath{\displaystyle}
i\in I\begin{cases}
z_i^{n_i}=z_0^{n_0}\\\\
\frac{1}{n_i}\frac{m_i}{z_i^{n_i-1}}=\frac{1}{n_0}\frac{m_0}{z_0^
{n_0-1}}
\end{cases},
\quad j\in J
\begin{cases}
-\frac{n_0}{n_j}\frac{m_j}{z_j^{n_j-1}}=z_0^{n_0}\\\\
\frac{1}{n_0}z_j^{n_j}=\frac{1}{n_0}\frac{m_0}{z_0^{n_0-1}}
\end{cases}
\quad,
\end{equation}
where $n_{\alpha}$ are arbitrary positive integers. For $J=\emptyset$ these systems furnish  the solutions  describing
the algebraic orbits of the Whitham hierarchy \cite{krich2}, while  the case $I=\emptyset$ includes the systems of string
equations considered by Takasaki \cite{tak1} and Wiegmann-Zabrodin
\cite{zab1}-\cite{zab5} in their applications of the dToda hierarchy. The discussion of our scheme for solving \eqref{3.1} requires
the consideration of the two cases $J=\emptyset$ and $J\neq\emptyset$ separately. In what follows we consider  solutions  with  only a finite number of  times $t_{\mu n}$  different from zero.

\subsection{ The case $J=\emptyset$}

The corresponding  system is given by
\begin{equation}\label{3.1a}\everymath{\displaystyle}
\begin{cases}
z_i^{n_i}=z_0^{n_0},\\\\
\frac{1}{n_i}\frac{m_i}{z_i^{n_i-1}}=\frac{1}{n_0}\frac{m_0}{z_0^
{n_0-1}},
\end{cases},\quad i=1,\ldots,M.
\end{equation}
The first group of equations \eqref{3.1a} is satisfied by setting
\begin{equation}\label{3.2}
z_i^{n_i}=z_0^{n_0}=E(p):=p^{n_0}+u_{{n_0}-2}p^{{n_0}-2}+\cdots+u_0+\sum_{i=1}^M\sum_{s=1}^{n_i}\frac{v_{i
s}} {(p-q_i)^s},
\end{equation}
and, obviously, appropriate branches of $z_{\mu}$ can be defined
which are compatible with the required asymptotic expansions
\eqref{1.1}. On the other hand, notice that  the remaining string
equations in \eqref{3.1a} can be rewritten as
\begin{equation}\label{3.3}
m_i=m_0\frac{\d z_0}{\d z_i},
\end{equation}
so that they are verified by taking
\begin{equation}\label{3.4}
m_{\alpha}=\frac{\partial S}{\partial z_{\alpha}},\quad \forall \alpha,
\end{equation}
for a given function $S(p,\bt)$, which means that all the
$S_{\alpha}$ are equal to $S$. Moreover, it is
straightforward to prove that the  expansions \eqref{1.2} are satisfied if we set
\begin{equation}\label{3.5}
S=\sum_{n=1}^{N_0}t_{0 n}(z_0^n)_{(0,
+)}+\sum_{j=1}^M\Big(\sum_{n=1}^{N_j}t_{j n}(z_j^n)_{(j, +)} -t_{j
0}\ln_j(p-q_j)\Big).
\end{equation}
In order to satisfy the hypothesis of Theorem 1,  the functions $z_0^{n_0}$ and
$m_0/z_0^{{n_0}-1}$
must be rational functions of $p$ with possible finite poles at the points $q_i$ only. In view of \eqref{3.2} this condition is verified by $z_0^n$. On the other hand, \eqref{3.1a} and \eqref{3.4}
imply that
\[
\frac{1}{{n_0}}\frac{m_0}{z_0^{{n_0}-1}}=\frac{\partial_p S}{\partial_p
E}.
\]
Therefore, the requirements of Theorem 1 are satisfied provided that
\begin{equation}\label{3.6}
\partial_p S(p_r)=0,
\end{equation}
where $p_r$ are the zeros of
\[
\partial_p E(p_r)=0.
\]
We observe that the number of zeros $p_r$ is
\[
M+n_0-1+\sum_{i=1}^M n_i,
\]
which equals the number of unknowns $\{q_i,u_0,\ldots,u_{{n_0}-2},
v_{i s}\}$. Equations \eqref{3.6} coincide with those formulated
by Krichever in \cite{krich2} for determining the algebraic orbits
of the Whitham hierarchy.

\vspace{0.3truecm}

\noindent \textbf{Example:}  $M=2$, $n_0=n_1=n_2=1$, $N_0=2$,
$N_1=N_2=1$\\

In this case \eqref{3.2} reads

$$z_{\alpha}=E(p)=p+\frac{v_1}{p-q_1}+\frac{v_2}{p-q_2},
\quad 0\leq\alpha\leq 2,$$
and \eqref{3.6} leads to

$$ -t_{1 0} - t_{2 0} +
2 t_{0 2} v_1 + 2 t_{0 2} v_2 = 0,$$

$$\begin{array}{l} q_1 t_{1 0} +
2 q_2 t_{1 0} + 2 q_1 t_{2 0} +
   q_2 t_{2 0} + t_{0 1} v_1 -
   4 q_2 t_{0 2} v_1 - t_{1 1} v_1 \\  \\
   +  t_{0 1} v_2 - 4 q_1 t_{0 2} v_2 -   t_{2 1} v_2 = 0,
\end{array}$$

$$\begin{array}{l}
-2 q_1 q_2 t_{1 0} - {q_2}^2 t_{1 0} -
   {q_1}^2 t_{2 0} - 2 q_1 q_2 t_{2 0} -
   2 q_2 t_{0 1} v_1 + 2 {q_2}^2 t_{0 2} v_1 \\  \\
   +   2 q_2 t_{1 1} v_1 - 2 q_1 t_{0 1} v_2 +
   2 {q_1}^2 t_{0 2} v_2 + 2 q_1 t_{2 1} v_2 =0,
\end{array}$$

$$ q_1 {q_2}^2 t_{1 0} +
{q_1}^2 q_2 t_{2 0} +
   {q_2}^2 t_{0 1} v_1 - {q_2}^2 t_{1 1} v_1 +
   {q_1}^2 t_{0 1} v_2 - {q_1}^2 t_{2 1} v_2 =
   0.$$
By solving this system  we  obtain

$$\everymath{\displaystyle}\begin{array}{rcl}
 z_{\alpha}=&E(p) = & p + \frac{2 t_{1 0} + t_{2 0}}
   {4 t_{0 2} \left( p + \dfrac{2 t_{0 1} t_{1 0} -
          2 t_{1 0} t_{1 1} + 2 t_{0 1} t_{2 0} -
          t_{1 1} t_{2 0} - t_{2 0} t_{2 1}}{4 t_{0 2}
          \left( t_{1 0} + t_{2 0} \right) } \right) } \\  \\
     &   &+  \frac{t_{2 0}}
   {4 t_{0 2} \left( p - \dfrac{-2 t_{0 1}
           \left( t_{1 0} + t_{2 0} \right)  + 2 t_{2 0} t_{2 1} +
         t_{1 0} \left(t_{1 1} + t_{2 1} \right) }{4 t_{0 2}
          \left( t_{1 0} + t_{2 0} \right) } \right)
          },\quad 0\leq\alpha \leq 2.\end{array}$$

\subsection{The case $J\neq\emptyset$}

Now we consider the system \eqref{3.1} for the generic case $J\neq\emptyset$. We
look for functions $m_{\alpha}$ of the form
\begin{equation}\label{3.2b}
m_{\alpha}(z,t)=\sum_{n=1}^{N_{\alpha}}nt_{\alpha
n}z_{\alpha}^{n-1}+\frac{t_{\alpha 0}} {z_{\alpha}}+
\sum_{n\geq2}\frac{v_{\alpha n}}{z_{\alpha}^n},\quad t_{0
0}=-\sum_{i=1}^Mt_{i 0},
\end{equation}
for arbitrary positive integers $N_{\alpha}$. In order to verify the hypothesis of Theorem 1 and the expansions
\eqref{1.1}, we set
\begin{equation}\label{3.4a}
\everymath{\displaystyle}
z_0^{{n_0}}=z_i^{n_i}=E_1(p):=p^{{n_0}}+u_{{n_0}-2}p^{{n_0}-2}+\cdots+u_0+
\sum_{l\in I}\sum_{n=1}^{n_l}\frac{a_{l n}}{(p-q_l)^n}+\sum_{k\in J}\sum_{n=1}^{n_{0k}}\frac{b_{k n}}{(p-q_k)^n},\quad \forall i\in I,
\end{equation}
\begin{equation}\label{3.4b}
z_j^{n_j}=E_2(p):=\sum_{n=0}^{n_{00}}c_n p^{n}+
\sum_{l\in I}\sum_{n=1}^{n_{0l}}\frac{\tilde{a}_{l n}}{(p-q_l)^n}+\sum_{k\in J}\sum_{n=1}^{n_k}\frac{\tilde{b}_{k n}}{(p-q_k)^n},\quad \forall j\in J,
\end{equation}
where $n_{00},n_{0 l},n_{0k}\;(l\in I,\;k\in J)$, the poles $q_i$ and the coefficients of $E_1$ and $E_2$ are to be determined.

By introducing the functions
\begin{equation}\label{3.3b}
{\cal M}_{\alpha}:=m_{\alpha}z_{\alpha},
\end{equation}
and taking \eqref{3.4a}-\eqref{3.4b} into account it follows that
the system of string equations \eqref{3.1} reduces to
\begin{equation}\label{3.4ab}
{\cal M}_0=\frac{{n_0}}{n_i}{\cal M}_i=-\frac{{n_0}}{n_j}{\cal M}_j=E_1(p)\,E_2(p)\quad \forall i\in I,
\; j\in J
\end{equation}
 On the other
hand, due to their rational character,  the functions ${\cal M}_{\alpha}$ can be written in terms of their
principal parts at the poles $q_{\beta}$
\[
{\cal M}_{\alpha}=\sum_{\beta=0}^M({\cal M}_{\alpha})_{(\beta, +)},
\]
and by taking \eqref{1.2} into account we get
\[
\everymath{\displaystyle}\begin{array}{l}
({\cal M}_0)_{(0,+)}=\sum_{n=1}^{N_0}nt_{0 n}(z_0^n)_{(0,+)}+t_{00},\quad t_{00}=-\sum_{i=1}^Mt_{i0},\\  \\
({\cal M}_i)_{(i,+)}=\sum_{n=1}^{N_i}nt_{i n}(z_i^n)_{(i,+)}.
\end{array}
\]
Therefore \eqref{3.4ab} is satisfied by
\begin{equation}\label{3.12}\everymath{\displaystyle}\begin{array}{l}
{\cal M}_0=\sum_{n=1}^{N_0}nt_{0n}(z_0^n)_{(0,+)}+t_{00}+
\sum_{i\in I} \frac{n_0}{n_i}\sum_{n=1}^{N_i}nt_{in}(z_i^n)_{(i,+)}
-
\sum_{j\in J}\frac{n_0}{n_j}\sum_{n=1}^{N_j}nt_{jn}(z_j^n)_{(j,+)},\quad t_{00}=-\sum_{j=1}^Mt_{j0}\\  \\
{\cal M}_i=\frac{n_i}{n}{\cal M}_0,\quad {\cal M}_j=-\frac{n_j}{n}{\cal M}_0,\quad \forall i\in I,
\; j\in J,
\end{array}\end{equation}
provided ${\cal M}_0$ verifies the  equation
\begin{equation}\label{3.14}
{\cal M}_0=E_1(p) E_2(p).
\end{equation}

At this point notice that from \eqref{3.4a},\eqref{3.4b} and \eqref{3.4ab} it follows that \eqref{3.14} is the only equation to be satisfied in order to solve the system of string equations \eqref{3.1}. Both sides of
\eqref{3.14} are rational functions of $p$ with finite poles  at
$\{q_1,\dots,q_M\}$ only, so that   \eqref{3.14} holds if
and only if  the principal parts of both members at
$\{q_0,q_1,\dots,q_M\}$ coincide. Now we have that:
\begin{itemize}
\item At $q_0=\infty$, the function ${\cal M}_0$ has a pole of order $N_0$, while $E_1(p)E_2(p)$ has a pole of order $n_{00}+n_0$, consequently \eqref{3.14} requires that  $n_{00}=N_0-n_0$, so that identifying the principal parts at $q_0$ represents $N_0+1$ equations.
\item At $q_i,\; (i\in I)$,  the function ${\cal M}_0$ has a pole of order $N_i$ and $E_1(p)E_2(p)$
has a pole of order $n_i+n_{0i}$. Hence  $n_{0 i}=N_i-n_i$ and
identifying the corresponding  principal parts leads to $N_i$
equations.
\item At $q_j,\; (j\in J)$,  the function ${\cal M}_0$ has a pole of order $N_j$ and $E_1(p)E_2(p)$
has a pole of order $n_{0j}+n_j$. Hence  $n_{0 j}=N_j-n_j$ and
identifying the corresponding  principal parts leads to $N_j$
equations.
\end{itemize}
Thus, Eq.\eqref{3.14} leads to
$N_0+\sum_{i=1}^MN_i+1$ equations. On the other hand we have $N_0+\sum_{i=1}^MN_i+M$
unknown coefficients given by
\begin{equation}\label{3.15}\begin{cases}
q_i,\quad  i=1,2,\dots,M,   \\
a_{i1},\ldots, a_{i n_i},\;\tilde{a}_{i1},\ldots, \tilde{a}_{i N_i-n_i},\quad i\in I ,\\
b_{j1},\ldots, b_{j N_j-n_j},\;\tilde{b}_{j1},\ldots, \tilde{b}_{jn_j},\quad j\in J ,  \\
u_0,\ldots,u_{n_0-2}  \\
c_0,\ldots,c_{N_0-n_0}
\end{cases}
\end{equation}
The additional $M-1$ equations required for determining these coefficients arise
by imposing the asymptotic behaviour \eqref{1.1}-\eqref{1.2} to $(z_{\alpha},m_{\alpha})$. Note that \eqref{3.4a} and \eqref{3.4b} imply that the functions $z_{\alpha}$ have the asymptotic form \eqref{1.1}. In what concerns the functions  $m_{\alpha}$, from the expression \eqref{3.12} for ${\cal M}_0$ it follows  that
\[
{\cal M}_0=\sum_{n=1}^{N_0}nt_{0n}z_0^n+t_{00}+\mathcal{O}\Big(\frac{1}{z_0}\Big),  \quad z_0\rightarrow\infty,
\]
so that $m_0$ satisfies an expansion of
the form \eqref{1.2}.  But in order for $m_i\; (i=1,2,\dots,M)$
to satisfy \eqref{1.2}  we must impose that
\begin{equation}\label{3.13}
\text{Res}(m_i,z_i=\infty)=t_{i0},\qquad i=1,2,\dots,M.
\end{equation}
However, let us see  that as a consequence of the string
equations \eqref{3.1} it follows that
\begin{equation}\label{3.14a}
\sum_{\alpha=0}^M \text{Res}(m_{\alpha},z_{\alpha}=\infty)=0,
\end{equation}
 and,
consequently, only $M-1$ of the equations \eqref{3.13} need to be
imposed.  Indeed,
we have
\[\everymath{\displaystyle}
2\pi i\sum_{\alpha=0}^M
\text{Res}(m_{\alpha},z_{\alpha}=\infty)=\sum_{\alpha=0}^M\oint_{\Gamma_{\alpha}}m_{\alpha}\d
z_{\alpha}=\sum_{\alpha=0}^M\oint_{\gamma_{\alpha}}m_{\alpha}\partial_p
z_{\alpha}\d p.
\]
On the other hand \eqref{3.4a},\eqref{3.4b} and \eqref{3.4ab} imply
\[
m_i\partial_p z_i=m_0\partial_p
z_0,\quad i\in I,
\]
\[
m_j\partial_p z_j=m_0\partial_p
z_0-\frac{1}{n_0}\partial_p (E_1(p) E_2(p)),\quad j\in J,
\]
so that

\[
2\pi i\sum_{\alpha=0}^M
 \text{Res}(m_{\alpha},z_{\alpha}=\infty)=\oint_{\gamma}m_0
\partial_p z_0\d
p=0,\quad \gamma:=\sum_{\alpha=0}^M \gamma_{\alpha},
\]
where we have taken into account that
\[
m_0\partial_p z_0=\frac{1}{n_0}E_2(p)\partial_p E_1(p),
\]
is a rational function of $p$ with finite poles at $q_i$ only, and
the fact that
\[
\gamma\sim 0 \quad \mbox{in}\quad
\mathbb{C}\setminus\{q_1,\dots,q_M\}.
\]

In this way we have a system of
\[
N_0+\sum_{i=1}^MN_i+M,
\]
equations to determine the same number of unknown coefficients.
Therefore, according to Theorem 1, this method furnishes solutions of the Whitham hierarchy.

\subsection{Examples}

\subsubsection*{1) $M=1,\; I=\emptyset,\; n_0=2,\; n_1=1,\; N_0=N_1=3$}
Note that in this case all the equations
come from \eqref{3.14}. We set
$$z_0^2=p^2+u_0+\frac{a_1}{p-q}+\frac{a_2}{(p-q)^2},\qquad z_1=\frac{b_1}{p-q}+c_0+c_1p.$$
From \eqref{3.14} one obtains the system
$$\everymath{\displaystyle}\begin{array}{lll}
p^3:      &   &3 t_{0 3} = c_1,\\  \\
p^2:      &   &2 t_{0 2} = c_0,\\   \\
p^1:      &   &t_{0 1} + \frac{9 t_{0 3} u_0}{2} = b_1+c_1 u_0,\\  \\
p^0:      &   &\frac{9 a_1 t_{0 3}}{2} - t_{1 0} +2 t_{0 2} u_0 =a_1 c_1 + b_1 q + c_0 u_0,\\  \\
(p-q)^{-3}:  &    &-6 {b_1}^2 t_{1 3} = a_2,\\  \\
(p-q)^{-2}:  &    &-2 {b_1}^2 \left( 2 t_{1 2} + 9 \left(c_0 +c_1 q \right)  t_{1 3} \right)  =\\  \\
             &    &a_1 b_1+ a_2 \left( c_0 + c_1 q \right),\\  \\
(p-q)^{-1}:  &    &-2 b_1\Big(t_{1 1} + 4\left(c_0 +c_1 q \right)t_{1 2} \\  \\
             &    &+ 9\left(c_0^2+ 2 c_0 c_1 q +c_1 b_1+c_1^2 q^2\right)  t_{1 3} \Big)  =  \\  \\
             &     &a_2 c_1 + a_1 \left(c_0 +c_1 q \right) +b_1 \left( q^2 +u_0 \right),
\end{array}$$
and by solving these equations we find
$$\everymath{\displaystyle}\begin{array}{l}
z_0^2 = p^2 \\  \\
      \quad- \frac{2\left( q t_{0 1} + t_{1 0} +6 t_{0 1} t_{0 3} t_{1 2} +
       36 t_{0 1} t_{0 2} t_{0 3} t_{1 3} + 54 q t_{0 1} {t_{0 3}}^2 t_{1 3} \right) }
       {3 t_{0 3} \left( q + 6 t_{0 3} t_{1 2} + 36 t_{0 2} t_{0 3} t_{1 3} +
       54 q {t_{0 3}}^2 t_{1 3} \right) }\\  \\
       \quad +  \frac{4 t_{1 0} \left(t_{1 2} + 6 t_{0 2} t_{1 3} +
       9 q t_{0 3} t_{1 3} \right) }{\left( p - q \right)
     \left( q + 54 q {t_{0 3}}^2 t_{1 3} +
       6 t_{0 3} \left(t_{1 2} + 6 t_{0 2} t_{1 3} \right)  \right) }\\  \\
       \quad -  \frac{6 {t_{1 0}}^2 t_{1 3}}
   {{\left( p - q \right) }^2 {\left( q + 54 q {t_{0 3}}^2 t_{1 3} +
         6 t_{0 3} \left( t_{1 2} + 6 t_{0 2} t_{1 3} \right)  \right) }^2},\\  \\  \\
z_1 =- \frac{t_{1 0}}
   {\left( p - q \right)  \left( q + 6 t_{0 3} t_{1 2} +
       36 t_{0 2} t_{0 3} t_{1 3} + 54 q {t_{0 3}}^2 t_{1 3} \right) }\\  \\
    \quad+2 t_{0 2} + 3 p t_{0 3},
\end{array}$$
where  $q$ is determined by the implicit equation
$$\everymath{\displaystyle}\begin{array}{l}
 -2 q t_{0 1} + 3 q^3 t_{0 3} -2 t_{1 0} + 6 q t_{0 3} t_{1 1} -12 t_{0 1} t_{0 3} t_{1 2} +
   24 q t_{0 2} t_{0 3} t_{1 2}  \\   \\
   + 54 q^2 {t_{0 3}}^2 t_{1 2}+36 {t_{0 3}}^2 t_{1 1} t_{1 2} +144 t_{0 2} {t_{0 3}}^2 {t_{1 2}}^2 +
   216 q {t_{0 3}}^3 {t_{1 2}}^2 \\  \\
   -72 t_{0 1} t_{0 2} t_{0 3} t_{1 3} + 72 q {t_{0 2}}^2 t_{0 3} t_{1 3} -
   108 q t_{0 1} {t_{0 3}}^2 t_{1 3} \\  \\
   + 324 q^2 t_{0 2} {t_{0 3}}^2 t_{1 3} +   324 q^3 {t_{0 3}}^3 t_{1 3}
   - 108 {t_{0 3}}^2 t_{1 0} t_{1 3} + 216 t_{0 2} {t_{0 3}}^2 t_{1 1} t_{1 3}\\  \\
   + 324 q {t_{0 3}}^3 t_{1 1} t_{1 3} + 1296 {t_{0 2}}^2 {t_{0 3}}^2 t_{1 2} t_{1 3}
   +  3888 q t_{0 2} {t_{0 3}}^3 t_{1 2} t_{1 3}\\  \\
   + 2916 q^2 {t_{0 3}}^4 t_{1 2} t_{1 3} +
   2592 {t_{0 2}}^3 {t_{0 3}}^2 {t_{1 3}}^2
   +11664 q {t_{0 2}}^2 {t_{0 3}}^3 {t_{1 3}}^2\\  \\
    + 17496 q^2 t_{0 2} {t_{0 3}}^4 {t_{1 3}}^2 +8748 q^3 {t_{0 3}}^5 {t_{1 3}}^2 = 0.
\end{array}$$
\subsubsection*{2) $M=2,\; I=\emptyset,\; n_0=n_1=n_2=1$, $N_0=N_1=2$, $N_2=1$}

 In this case there are three punctures $\{q_0=\infty,q_1,q_2\}$ and
we have to impose equation \eqref{3.13} for $i=1$.
We take
$$z_0=p+\frac{a_1}{p-q_1},\qquad z_1=z_2=\frac{b_1}{p-q_1}+\frac{b_2}{p-q_2}+c_0+c_1p.$$
Then, by identifying powers of $p$, $(p-q_1)^{-1}$ and
$(p-q_2)^{-1}$ in \eqref{3.14} the following system of equations
arises
$$\everymath{\displaystyle}\begin{array}{lll}
p^2:   &    &2 t_{0 2} = c_1,\\  \\
p^1:   &    &t_{0 1} = c_0,\\  \\
p^0:   &    &4 a_1 t_{0 2} - t_{1 0} - t_{2 0} =b_1 + b_2 + a_1 c_1,\\  \\
(p-q_1)^{-2}:   &    &-2 {b_1} t_{1 2} = a_1,\\  \\
(p-q_1)^{-1}:   &    &-b_1 t_{1 1} -4 b_1 \left(c_0 +c_1 q_1+ \frac{b_2}{q_1 - q_2} \right)  t_{1 2} =\\  \\
                &    &  b_1 q_1 + a_1 \left(c_0 +c_1 q_1 +\frac{b_2}{q_1 -q_2} \right),\\  \\
(p-q_2)^{-1}:   &    &-t_{2 1} = q_2 - \frac{a_1}{q_1 - q_2}.
\end{array}$$
Moreover, by taking \eqref{3.12} into account, from \eqref{3.13}
we get
$$\everymath{\displaystyle}\begin{array}{l}
-q_1 t_{0 1} -   2 \left( 2 a_1 + {q_1}^2 \right)  t_{0 2} -
   \left( c_0 + c_1 q_1 +  \frac{b_2}{q_1 - q_2} \right)  t_{1 1} \\  \\
-2 \left( 2 b_1 \left( c_1 - \frac{b_2}{{\left(q_1 - q_2 \right)
}^2} \right)  +
      {\left( c_0 + c_1 q_1 + \frac{b_2}{q_1 -q_2} \right) }^2 \right)  t_{1 2}\\  \\
       +   t_{2 0} + \frac{b_2 t_{2 1}}{q_1 -q_2} = 0.
\end{array}$$
These equations lead to
$$\everymath{\displaystyle}\begin{array}{lll}
z_0&=&p -\frac{1}{2\left(1 +4 t_{0 2} t_{1 2} \right)  \left( p
-q_1\right)}\Big(
      {r}^2 \left( 1 +4 t_{0 2} t_{1 2} \right) -
     2 t_{1 2} \left( t_{1 0} + t_{2 0} \right)\\  \\
   & &\qquad  + r \left(t_{1 1}+ 2 t_{0 1} t_{1 2} - t_{2 1} -4 t_{0 2} t_{1 2} t_{2 1} \right)\Big),
     \\  \\  \\
z_1&=& \frac{1}{4 t_{1 2} \left( 1 + 4 t_{0 2} t_{1 2} \right)\left(
p - q_1\right)}
       \Big({r}^2 \left( 1 + 4 t_{0 2} t_{1 2} \right)  -
     2 t_{1 2} \left(t_{1 0} +t_{2 0} \right)  \\  \\
   & &  + r \left( t_{1 1} + 2 t_{0 1} t_{1 2} - t_{2 1} -4 t_{0 2} t_{1 2} t_{2 1} \right) \Big)\\  \\
   & & +\frac{1}{4 t_{1 2} \left( p - q_2\right)}
       \Big(-{r}^2 \left( 1 + 4 t_{0 2} t_{1 2} \right)  -
     2 t_{1 2} \left(t_{1 0} +t_{2 0} \right)  \\  \\
   & &  + r \left(- t_{1 1} - 2 t_{0 1} t_{1 2} + t_{2 1} +4 t_{0 2} t_{1 2} t_{2 1} \right) \Big)\\  \\
   & &+t_{0 1}+2 p  t_{0 2},
\end{array}$$
where
$$\everymath{\displaystyle}\begin{array}{lll}
q_1&=&\frac{1} {2 r  \left(1+ 4 t_{0 2} t_{1 2} \right) }\Big(
      {r}^2 \left( 1 + 4 t_{0 2} t_{1 2}\right)  +2 t_{1 2} \left(t_{1 0} + t_{2 0} \right)\\  \\
   & & -r \left( t_{1 1} + 2 t_{0 1} t_{1 2} + t_{2 1} +
       4 t_{0 2} t_{1 2} t_{2 1} \right) \Big)
      ,\\  \\
q_2&=&- \frac{1}{2 r \left(1+ 4 t_{0 2} t_{1 2} \right) }\Big(
       {r}^2 \left( 1 +4 t_{0 2} t_{1 2} \right)  -2 t_{1 2} \left( t_{1 0} + t_{2 0} \right)\\  \\
   & & + r \left( t_{1 1} + 2 t_{0 1} t_{1 2} + t_{2 1} +
         4 t_{0 2} t_{1 2} t_{2 1} \right) \Big),
\end{array}$$
and $r$ is determined by the equation:
$$\everymath{\displaystyle}\begin{array}{lll}
 3 {r}^4 {\left( 1 + 4 t_{0 2} t_{1 2}\right) }^2 -
   4 {t_{1 2}}^2 {\left( t_{1 0} + t_{2 0} \right) }^2 \\  \\
   +4 {r}^3 \left( 1 + 4 t_{0 2} t_{1 2} \right)
    \left( t_{1 1} + 2 t_{0 1} t_{1 2} -
      \left( 1 + 4 t_{0 2} t_{1 2} \right)  t_{2 1} \right)  \\  \\
   +{r}^2 \Big( {t_{1 1}}^2 + 4 {t_{0 1}}^2 {t_{1 2}}^2 +
      4 t_{1 0} t_{1 2} \left( 1 + 4 t_{0 2} t_{1 2} \right)  -
      4 t_{1 2} t_{2 0} - 16 t_{0 2} {t_{1 2}}^2 t_{2 0} \\  \\
      -      4 t_{0 1} t_{1 2} t_{2 1} -
      16 t_{0 1} t_{0 2} {t_{1 2}}^2 t_{2 1} + {t_{2 1}}^2 +
      8 t_{0 2} t_{1 2} {t_{2 1}}^2 +
      16 {t_{0 2}}^2 {t_{1 2}}^2 {t_{2 1}}^2 \\  \\
      +      2 t_{1 1} \left( 2 t_{0 1} t_{1 2} -
         \left( 1 + 4 t_{0 2} t_{1 2} \right)  t_{2 1} \right)  \Big)  = 0.
\end{array}$$
\subsubsection*{3) $M=2,\; I=\{1\},\; J=\{2\},\; n_0=n_1=n_2=N_0=1,\;N_1=N_2=2$.}

In this case we take
$$z_0=z_1=E_1(p)=p + \frac{v_{1 1}}{p - q_1} + \frac{v_{2 1}}{p -q_2},
\quad z_2=E_2(p)=\frac{w_{2 1}}{p - q_2}+\frac{w_{1 1}}{p - q_1} +c_0.$$
By equating the coefficients of $p^1$, $p^0$, $(p-q_i)^{-j}$,
$i,j=1,2$ in  \eqref{3.14}  one finds
$$\everymath{\displaystyle}\begin{array}{lcl}
p^1:&  &t_{0 1} = c_0,\\  \\
p^0:&  & -t_{1 0} - t_{2 0} = w_{1 1} + w_{2 1},\\  \\
(p-q_1)^{-2}:&   &2 t_{1 2} {v_{1 1}}^2 = v_{1 1} w_{1 1},\\  \\
(p-q_1)^{-1}:&   &v_{1 1} \left( t_{1 1} + 4 t_{1 2}
      \left( q_1 + \frac{v_{2 1}}{q_1 - q_2} \right)  \right)\\  \\
             &   &  =  \left( q_1 + \frac{v_{2 1}}{q_1 - q_2} \right)  w_{1 1} +
   v_{1 1} \left( c_0 + \frac{w_{2 1}}{q_1 - q_2} \right),\\  \\
(p-q_2)^{-2}:&   &-2 t_{2 2} {w_{2 1}}^2 = v_{2 1} w_{2 1},\\  \\
(p-q_2)^{-1}:&   &-\left( t_{2 1} w_{2 1} \right) -
   4 t_{2 2} \left( c_0 + \frac{w_{1 1}}{-q_1 + q_2} \right)
       w_{2 1}\\  \\
             &   &= c_0 v_{2 1} +
   \frac{v_{2 1} w_{1 1} + \left( -\left( q_1 q_2 \right)  +
         {q_2}^2 + v_{1 1} \right)  w_{2 1}}{-q_1 + q_2},
\end{array}$$
and \eqref{3.13} leads to the equation
$$\everymath{\displaystyle}\begin{array}{c}- q_2 t_{0 1}  + t_{1 0} -
   \frac{t_{1 1} v_{1 1}}{-q_1 + q_2} -
   2 t_{1 2} \left( \frac{{v_{1 1}}^2}
       {{\left( -q_1 + q_2 \right) }^2} +
      \frac{2 v_{1 1} \left( q_1 +
           \frac{v_{2 1}}{q_1 - q_2} \right) }{-q_1 + q_2}
      \right) \\  \\
       - t_{2 1} \left( c_0 +
      \frac{w_{1 1}}{-q_1 + q_2} \right)  -
   2 t_{2 2} \left( {\left( c_0 +
          \frac{w_{1 1}}{-q_1 + q_2} \right) }^2 -
      \frac{2 w_{1 1} w_{2 1}}{{\left( q_1 - q_2 \right) }^2} \right)  =
   0.\end{array}$$
By solving these equations one obtains
$$\everymath{\displaystyle}\begin{array}{lcl}
E_1(p)= p \\   \\
      \quad - \frac{2{r_1}^2t_{1 2} +
     \left(t_{1 0} + t_{2 0} \right)\left( 1 + 4t_{1 2} t_{2 2} \right)  -
     r_1\left(t_{0 1} - t_{1 1} + 2t_{1 2}t_{2 1} +
        4t_{0 1}t_{1 2}t_{2 2} \right) }{4\left( p - q_1 \right)
     t_{1 2}\left( 1 + 4t_{1 2}t_{2 2} \right) } \\  \\
       \quad - \frac{t_{2 2}\left(2 {r_1}^2t_{1 2} -
       \left(t_{1 0} + t_{2 0} \right) \left( 1 + 4t_{1 2}t_{2 2} \right)  -
       r_1\left( t_{0 1} - t_{1 1} + 2t_{1 2}t_{2 1} +
          4t_{0 1}t_{1 2}t_{2 2} \right)  \right) }{\left( p -
          q_2 \right) \left( 1 + 4t_{1 2}t_{2 2} \right) },
\end{array}$$
$$\everymath{\displaystyle}\begin{array}{l}
E_2(p) = t_{0 1} \\  \\
      \quad -\frac{2 {r_1}^2t_{1 2} +
     \left( t_{1 0} + t_{2 0} \right)
      \left( 1 + 4 t_{1 2}t_{2 2} \right)  -
     r_1\left(t_{0 1} - t_{1 1} + 2t_{1 2}t_{2 1} +
        4t_{0 1}t_{1 2}t_{2 2} \right) }{2\left( p - q_1 \right)
     \left( 1 + 4 t_{1 2}t_{2 2} \right) }\\  \\
     \quad  + \frac{2{r_1}^2t_{1 2} -
     \left( t_{1 0} + t_{2 0} \right)
      \left( 1 + 4t_{1 2}t_{2 2} \right)  -
     r_1\left( t_{0 1} - t_{1 1} + 2t_{1 2}t_{2 1} +
        4t_{0 1}t_{1 2}t_{2 2} \right) }{2\left( p - q_2 \right)
     \left( 1+4t_{1 2}t_{2 2} \right) },
\end{array}$$
where
$$q_1= \frac{r_1}{2}  - \frac{\left( t_{1 0} + t_{2 0}
\right)
        \left( 1 + 4 t_{1 2} t_{2 2} \right)  +
       r_1 \left( t_{1 1} + 2 t_{1 2} t_{2 1} +
          t_{0 1} \left( -1 + 4 t_{1 2} t_{2 2} \right)  \right) }{4
       r_1 t_{1 2}},$$
$$q_2=-\frac{r_1}{2}  - \frac{\left( t_{1 0} + t_{2 0}
\right)
        \left( 1 + 4 t_{1 2} t_{2 2} \right)  +
       r_1 \left( t_{1 1} + 2 t_{1 2} t_{2 1} +
          t_{0 1} \left( -1 + 4 t_{1 2} t_{2 2} \right)  \right) }{4
       r_1 t_{1 2}},$$
and $r_1$ satisfies the equation:
$$\everymath{\displaystyle}\begin{array}{c}
 -12 {r_1}^4 {t_{1 2}}^2 +
   {\left( t_{1 0} + t_{2 0} \right) }^2
      {\left( 1 + 4 t_{1 2} t_{2 2} \right) }^2\\  \\
       + 8 {r_1}^3 t_{1 2} \left( t_{0 1} - t_{1 1} +
      2 t_{1 2} t_{2 1} + 4 t_{0 1} t_{1 2} t_{2 2} \right) \\  \\
       -{r_1}^2\Big( {t_{1 1}}^2 -
        4t_{1 1}t_{1 2}t_{2 1} -
        2t_{0 1}\left(t_{1 1} - 2t_{1 2}t_{2 1} \right)\left( 1 + 4t_{1 2}t_{2 2} \right) +
         {\left( t_{0 1} + 4t_{0 1}t_{1 2}t_{2 2} \right) }^2 \\  \\
          +        4 t_{1 2} \left( -t_{1 0} + t_{2 0} + t_{1 2} {t_{2 1}}^2 -
           4 t_{1 0} t_{1 2} t_{2 2} +
           4 t_{1 2} t_{2 0} t_{2 2} \right)  \Big)  = 0.
\end{array}$$
\subsection{$S$-functions}

According  to the identities
\[
\partial_p S_{\alpha}={\cal M}_{\alpha}\partial_p \log z_{\alpha},\quad {\cal M}_{\alpha}=m_{\alpha}z_{\alpha},
\]
it follows at once from \eqref{3.4a},\eqref{3.4b} and \eqref{3.4ab} that the functions $\partial_p S_{\alpha}$
are rational functions of $p$ with finite poles at the points $q_i\; (i=1,\ldots,M)$ only. Thus we may decompose the functions $\partial_p S_{\alpha}$ into their principal parts
\begin{equation}\label{t0}
\partial_p S_{\alpha}=\sum_{\beta}\Big(\partial_p S_{\alpha}\Big)_{(\beta,+)},
\end{equation}
and, in view of the assymptotic behaviour \eqref{2aa}, we may  write
\begin{equation}\label{t00}
\Big(\partial_p S_{\alpha}\Big)_{(\alpha,+)}=\partial_p R_{\alpha}.
\end{equation}
where
\begin{equation}\label{t000}
R_{\alpha}:=\sum_{n\geq 1}(z_{\alpha}^n)_{(\alpha,+)}t_{\alpha
n}-(1-\delta_{\alpha 0})t_{\alpha 0}\log_{\alpha} (p-q_{\alpha}).
\end{equation}
Furthermore, from  \eqref{3.4a}, \eqref{3.4b} and \eqref{3.4ab}  we obtain
\begin{equation}\label{t1}\begin{cases}
\partial_p S_i=\partial_p S_0,\quad \forall i\in I,\\
\partial_p S_j=\partial_p S_0-\frac{1}{n_0}\partial_p (E_1 E_2),\quad \forall j\in J,
\end{cases}
\end{equation}
which leads  to
\begin{equation}\label{ds}\everymath{\displaystyle}
S_{\alpha}=
\begin{cases}
\sum_{\beta} R_{\beta}+\frac{1}{n_0}\sum_{j\in J} (E_1\,E_2)_{(j,+)},\quad \alpha\in\{0\}\cup I,\\\\
\sum_{\beta} R_{\beta}-\frac{1}{n_0}(E_1\,E_2)_{(0,+)}-\frac{1}{n_0}\sum_{i\in I} (E_1\,E_2)_{(i,+)},\quad \alpha\in J.
\end{cases}
\end{equation}
In principle \eqref{t1}  implies the expressions \eqref{ds} plus  additional
$p$-independent terms $w_{\mu}(t)$. However these terms can be removed by using  \eqref{2aaa} and \eqref{2aa}. Indeed, the assymptotic behavior \eqref{2aa} for $S_0$ requires $w_0=0$. On the other hand \eqref{2aaa} says that
\[
\d S_i-\d S_0=m_i\d z_i-m_0\d z_0,
\]
so that by using the string equations \eqref{3.1} we deduce
\[
\d w_i=
\begin{cases}
\d S_i-\d S_0 =0,\quad i\in I,\\\\
\d S_i-\d S_0+\frac{1}{n_0}\d (E_1\,E_2)=0,\quad i\in J.
\end{cases}
\]
\subsection{$\tau$-functions}

\begin{teh}
The $\tau$-function for the solutions of the Whitham hierarchy associated with the  class of string equations \eqref{3.1} is given by
\begin{align}\label{fe}\everymath{\displaystyle}
\nonumber 2\log\tau&=\sum_{\alpha}\frac{1}{2\pi i}\oint_{\Gamma_{\alpha}}
(\sum_{n\geq 1}z_{\alpha}^n t_{\alpha n})\,m_{\alpha}\d z_{\alpha}
-\sum_{j\in J}\frac{1}{4\pi i\,n_j}
\oint_{\Gamma_j}z_j m_j^2 \d z_j
+\sum_i t_{i0} v_{i1}\\\\
\nonumber &=\sum_{\alpha}\sum_{n\geq 0}t_{\alpha n}v_{\alpha
n+1}-\sum_{j\in J}\frac{1}{n_j}\sum_{n\geq 1}n\, t_{j n}v_{j n+1}
-\sum_{j\in J}\frac{t_{j0}^2}{2\,n_j}
\end{align}
\end{teh}
\begin{proof}

Our strategy to prove \eqref{fe} is to start from the free-energy function for the algebraic orbits of the Whitham  hierarchy \cite{krich2}
\begin{equation}\label{t6}
F_0:=\sum_{\alpha}\frac{1}{4\pi i}\oint_{\Gamma_{\alpha}}
(\sum_{n\geq 1}z_{\alpha}^n t_{\alpha n})\,m_{\alpha}\d z_{\alpha}
+\frac{1}{2}\sum_i t_{i0}v_{i1},
\end{equation}
and determine the  appropriate modifications to get the free-energy function for the solutions of the class of string equations \eqref{3.1}.

By differentiating $F_0$ with respect to  $t_{ln},\,n\geq 1$ we get
\begin{equation}\label{t7}
\partial_{l,n} F_0=\frac{1}{4\pi i}\oint_{\Gamma_l}z_l^n m_l\d z_l+
\sum_{\alpha}\frac{1}{4\pi i}\oint_{\gamma_{\alpha}}(\sum_{m\geq 1}z_{\alpha}^m t_{\alpha m})\partial_p \Big((z_l^n)_{(l,+)}\Big)\d p+\frac{1}{2}\sum_i t_{i0}\partial_{l,n} v_{i1},
\end{equation}
and arguing as in the derivation of \eqref{cross}, we find that
\[
\oint_{\gamma_{\alpha}}(\sum_{m\geq 1}z_{\alpha}^m t_{\alpha m})\partial_p \Big((z_l^n)_{(l,+)}\Big)\d p=
\oint_{\gamma_l}z_l^n \partial_p\Big(\sum_{m\geq 1}(z_{\alpha}^m)_{(\alpha,+)} t_{\alpha m}\Big)\d p.
\]
On the other hand, for $i\neq l$ we have
\[
\partial_{l,n} S_i=(z_l^n)_{(l,+)}=-\partial_{l,n} v_{i1}+\mathcal{O}\Big(\frac{1}{z_i}\Big),\quad p\rightarrow q_i,
\]
so that
\begin{align*}\everymath{\displaystyle}
\partial_{l,n} v_{i1}&=-(z_l^n)_{(l,+)}(q_i)=\frac{1}{2\pi i}\oint_{\gamma_i}\frac{(z_l^n)_{(l,+)}}{p-q_i}\d p\\
&=-\frac{1}{2\pi i}\oint_{\gamma_l}(z_l^n)_{(l,+)}\partial_p \log (p-q_i)\d p=
-\frac{1}{2\pi i}\oint_{\gamma_l}z_l^n\partial_p \log (p-q_i)\d p
,\quad
i\neq l,
\end{align*}
and the same expression turns out to hold for $i=l$. Hence \eqref{t7}
can be rewritten as
\begin{align}\label{t8}\everymath{\displaystyle}
\nonumber \partial_{l,n} F_0&=\frac{1}{2}v_{ln+1}+
\frac{1}{4\pi i}\oint_{\gamma_l}z_l^n\sum_{\alpha}\partial_p \Big(
\sum_{m\geq 1}(z_{\alpha}^m)_{(\alpha,+)} t_{\alpha m}-(1-\delta_{\alpha 0})t_{\alpha 0}\log(p-q_{\alpha})\Big)\d p\\\nonumber\\
&=v_{ln+1}+
\frac{1}{4\pi i}\oint_{\gamma_l}z_l^n\Big(\sum_{\alpha}\partial_p R_{\alpha}-\partial_p S_l\Big)\d p\\\nonumber\\
\nonumber &=
v_{ln+1}+\sum_{\alpha}
\frac{1}{4\pi i}\oint_{\gamma_l}(z_l^n)_{(l,+)}\Big(\partial_p R_{\alpha}-(\partial_p S_l)_{(\alpha,+)}\Big)\d p.
\end{align}
Furthermore, we have that
\begin{align*}\everymath{\displaystyle}
\frac{1}{4\pi i}\oint_{\gamma_l}(z_l^n)_{(l,+)}&\Big(\partial_p R_{\alpha}-(\partial_p S_l)_{(\alpha,+)}\Big)\d p=
-\frac{1}{4\pi i}\oint_{\gamma_{\alpha}}(z_l^n)_{(l,+)}\Big(\partial_p R_{\alpha}-(\partial_p S_l)_{(\alpha,+)}\Big)\d p\\\\
&=-\frac{1}{4\pi i}\oint_{\gamma_{\alpha}}(z_l^n)_{(l,+)}\Big(\partial_p S_{\alpha}-\partial_p S_l\Big)\d p,
\end{align*}
so that from \eqref{t1} and by taking into account that
\[
\oint_{\gamma_0}(z_l^n)_{(l,+)}\partial_p (E_1\,E_2)\d p=
-\sum_{i=1}^M\oint_{\gamma_i}(z_l^n)_{(l,+)}\partial_p (E_1\,E_2)\d p,
\]
we get
\begin{align}\label{t9}\everymath{\displaystyle}
\nonumber &\partial_{l,n} F_0=v_{ln+1}+\frac{1}{n_0}\sum_{j\in J}\frac{1}{4\pi i}
\oint_{\gamma_j}(z_l^n)_{(l,+)}\partial_p (E_1\,E_2)\d p\\\\
\nonumber&=v_{ln+1}+\sum_{j\in J}\frac{1}{4\pi i\, n_j}
\oint_{\Gamma_j}z_j m_j \partial_{l,n}m_j \d z_j
=v_{ln+1}+\partial_{l,n}\Big(\sum_{j\in J}\frac{1}{8\pi i\,n_j}
\oint_{\Gamma_j}z_j m_j^2 \d z_j\Big),
\end{align}
which shows that
\[
\partial_{l,n} \log\tau =v_{ln+1}.
\]

By a similar procedure one finds
\[
\partial_{0,n} \log\tau =v_{0n+1},\quad n\geq 1.
\]
Nevertheless, proving that
\begin{equation}\label{l0}
\partial_{l,0} \log\tau=v_{l1},\quad l=1,\ldots,M.
\end{equation}
requires a more involved analysis. Firstly
we differentiate $F_0$ with respect to $t_{l0}$
\begin{equation}\label{t11}
\partial_{l,0} F_0=
\sum_{\alpha}\frac{1}{4\pi i}\oint_{\gamma_{\alpha}}(\sum_{m\geq 1}z_{\alpha}^m t_{\alpha m})\partial_p \Big(-\log(p-q_l)\Big)\d p+\frac{1}{2}\sum_i t_{i0}\partial_{l,0} v_{i1}+\frac{1}{2}v_{l1},
\end{equation}
and use the following relations
\[
\oint_{\gamma_{\alpha}}(\sum_{m\geq 1}z_{\alpha}^m t_{\alpha m})
\frac{1}{p-q_l}\d p=-
\oint_{\gamma_l}\Big(\sum_{m\geq 1}(z_{\alpha}^m)_{(\alpha,+)} t_{\alpha m}\Big)\frac{1}{p-q_l}\d p,\quad \alpha\neq l,
\]
\[
\oint_{\gamma_l}(\sum_{m\geq 1}z_l^m t_{l m})
\frac{1}{p-q_l}\d p=-2\pi i \lim_{p\rightarrow q_l}
\Big(\sum_{m\geq 1}(z_l^m-(z_l^m)_{(l,+)}) t_{l m}\Big),
\]
\begin{align*}
&\sum_{i\neq l}t_{i0}\partial_{l0}v_{i1}=\sum_{i\neq
l}t_{i0}\log_l(q_i-q_l)-\sum_{i> l}t_{i0}\log_{li}(-1)\\
&=\sum_{i\neq l}t_{i0}\log_i(q_l-q_i)-\sum_{i<l}t_{i0}\log_{il}(-1),
\end{align*}
\[
\partial_{l,0} v_{l1}=\lim_{p\rightarrow q_l}\log(z_l(p-q_l)).
\]
Then \eqref{t11} becomes
\begin{align}\label{t12}\everymath{\displaystyle}
\nonumber &\partial_{l,0} F_0=\frac{1}{2}v_{l1}+ \frac{1}{4\pi
i}\oint_{\gamma_l}\frac{\d p}{p-q_l}\sum_{\alpha\neq l}\Big(
\sum_{m\geq 1}(z_{\alpha}^m)_{(\alpha,+)} t_{\alpha m}-(1-\delta_{\alpha 0})t_{\alpha 0}\log_{\alpha}(p-q_{\alpha})\Big)\\
\\
\nonumber &+\frac{1}{2}\lim_{p\rightarrow q_l}\Big(\sum_{m\geq
1}(z_l^m-(z_{l}^m)_{(l,+)}) t_{l m}+t_{l0}\log
z_l+t_{l0}\log_l(p-q_l))\Big)-\frac{1}{2}\sum_{i<l}t_{i0}\log_{il}(-1).
\end{align}
Furthermore, from the asymptotic expansion \eqref{2aa} of $S_l$ we have that
\[
v_l=\lim_{p\rightarrow q_l}\Big(\sum_{m\geq 1}z_l^m t_{l m}+t_{l0}
\log z_l-S_l\Big),
\]
which allows us to rewrite \eqref{t12} in the form
\begin{align}\label{t13}\everymath{\displaystyle}
\nonumber \partial_{l,0} F_0&=v_{l1}+
\frac{1}{4\pi i}\oint_{\gamma_l}\frac{\d p}{p-q_l}\sum_{\alpha\neq l}
R_{\alpha}+\frac{1}{2}\lim_{p\rightarrow q_l}\Big(S_l-R_l\Big)\\\\
\nonumber &=v_{l1}+\frac{1}{4\pi i}\oint_{\gamma_l}\frac{\d
p}{p-q_l}\Big(\sum_{\alpha} R_{\alpha}-S_l\Big).
\end{align}
Now by using \eqref{ds} it is straightforward to see that
\begin{equation}\label{int}\everymath{\displaystyle}
\frac{1}{4\pi i}\oint_{\gamma_l}\frac{\d p}{p-q_l}\Big(\sum_{\alpha}
R_{\alpha}-S_l\Big)=
\frac{1}{n_0}\sum_{j\in J}\frac{1}{4\pi i}
\oint_{\gamma_j}\frac{\d p}{p-q_l} E_1\,E_2
=\sum_{j\in J}\frac{1}{4\pi i\,n_j}
\oint_{\Gamma_j}z_j m_j \partial_{l,0}m_j \d z_j,
\end{equation}
which shows that \eqref{t13} is equivalent to \eqref{l0}.
\end{proof}
\begin{figure}\label{fig 2}
\begin{center}
\psfrag{p}{$p$-plane}\psfrag{z}{$z$-plane}\psfrag{d}{$D$}
\psfrag{G}{$\Gamma$}\psfrag{g}{$\gamma$}\psfrag{q}{$q$}
\includegraphics[width=13cm]{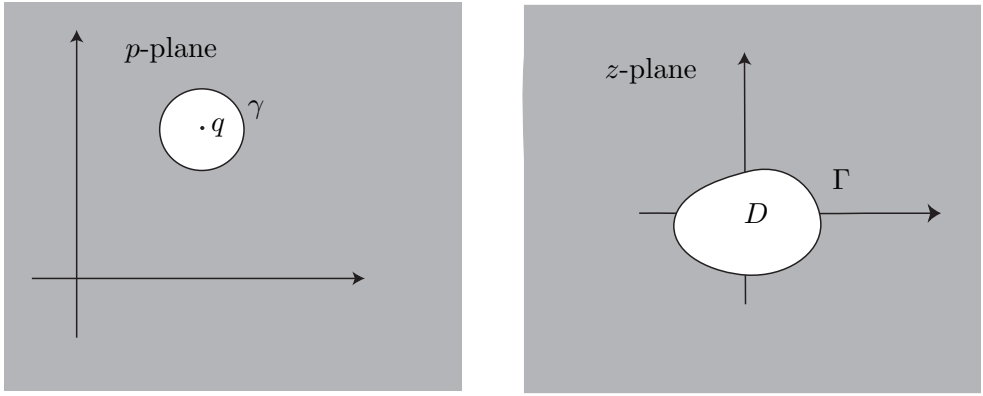}
\end{center}\caption{Conformal map $z=z(p)$}
\end{figure}

\subsection{Conformal maps dynamics}

 We will  outline  how our scheme  applies for characterizing dToda dynamics of conformal maps,
and, in particular, how
\eqref{fe} gives rise to the expression of the $\tau$-function of
analytic curves found in \cite{zab1}.

Given a simply-connected domain $D$ bounded by a closed path $\Gamma$ in the $z$-plane, there exist \cite{hen} a unique circle $\gamma$ in the $p$-plane and a unique conformal map $z=z(p)$ satisfying
\begin{equation}\label{c1}
z(p)=p+\sum_{n=1}^{\infty}\frac{d_n}{p^n}, \quad z\rightarrow\infty,
\end{equation}
such that $z=z(p)$ transforms the exterior of $\gamma$ into the exterior of $\Gamma$ (see figure 2). Note that the conformal map used by Wiegmann-Zabrodin  \cite{zab1} is given by  $z(r\,p+q)$, where $q$ and $r$ are the center and the radius of $\gamma$, respectively.

Let us define the function
\begin{equation}\label{c2}
\bar{z}(p)=\overline{z(\mathcal{I}_{\gamma}(p))},
\end{equation}
where $\mathcal{I}_{\gamma}$ denotes the inversion with respect to the circle $\gamma$
\[
\mathcal{I}_{\gamma}(p):=q+\frac{r^2}{\bar{p}-\bar{q}}.
\]
It is clear that
\[
\bar{z}(p)=\overline{z(p)},\quad \forall p\in\gamma.
\]
If $\Gamma$ is asssumed to be an analytic curve, then it can be described by an equation of the form
\begin{equation}\label{c3}
\bar{z}=\mathcal{S}(z),
\end{equation}
where $\mathcal{S}(z):=\bar{z}(p(z))$ (the Schwarz function) is analytic in a neigborhood of $\gamma$. Thus, if $\Gamma$ encircles the origin, $\mathcal{S}(z)$ can be expanded as
\begin{equation}\label{c4}
\mathcal{S}(z)=\sum_{n\geq 1}n\,t_n\,z^{n-1}+\frac{t_0}{z}+\sum_{n\geq 1}\frac{v_n(t)}{z^{n+1}},
\end{equation}
where the coefficients $t_n\,(n\geq 0)$, the exterior harmonic moments of $\Gamma$,  determine the curve $\Gamma$ and the conformal map \eqref{c1}. Note , in particular,
that the coefficient $t_0$
\[
t_0=\frac{1}{2\pi i}\oint_{\Gamma}\bar{z}\d z=\frac{1}{\pi}\int_{D}\d x \d y,
\]
represents the area of $D$.

In this way, by considering the harmonic moments as independent complex parameters, if we  define
\begin{align*}
& q_1=q,\quad z_0(p)=z(p),\quad z_1(p)=\bar{z}(p),\\
& t_{0n}:=t_n, \quad  t_{1n}=-\bar{t}_n,\quad n\geq 0,\\
& m_0=\mathcal{S}(z_0),\quad m_1=-z(p(z_1)),
\end{align*}
we obtain a solution of the system of string equations
\begin{equation}\label{c5}
-m_1=z_0,\quad z_1=m_0.
\end{equation}
Furthermore, by taking into account that
\[
v_{1n+1}=-\bar{v}_{0n+1}=-\bar{v}_n,\quad n\geq 1,
\]
and the identity
\begin{align*}
&t_{0}^2+2\,\sum_{n\geq 1}n\,t_n\,v_n= \frac{1}{2\pi i}\oint_{\Gamma}z
\mathcal{S}(z)^2\d z=\frac{1}{2\pi i}\oint_{\Gamma}z \bar{z}^2\d z\\\\
&=\frac{1}{\pi}\int_{D}|z|^2\d x \d y=\bar{t}_{0}^2+2\,\sum_{n\geq
1}n\,\bar{t}_n\,\bar{v}_n,
\end{align*}
we see that \eqref{fe} reduces to
\begin{equation}\label{c7}
2\log\tau=\frac{1}{2}\sum_{n\geq 1}(2-n)(t_n\,v_n+\bar{t}_n
\,\bar{v}_n)+t_0\,v_0-\frac{t_0^2}{2},
\end{equation}
where $v_0:=-v_{11}$, which is the expression  for the $\tau$-function associated to analytic curves obtained in \cite{zab1}. Notice that \eqref{c5} is the simplest nontrivial case
($I=\emptyset,\; J=\{1\},\; n_0=n_1=1$) of the class of string equations  \eqref{3.1} .

\subsection{Symmetry constraints}

As we proved above,  solutions $(z_{\alpha},m_{\alpha})$ of systems of string equations \eqref{2.1}
are invariant under the symmetries
\[
\bV_{rs}=(P_0^{r+1}\,Q_0^{s+1},\ldots,P_M^{r+1}\, Q_M^{s+1}),\quad
r\geq -1,\,s\geq 0.
\]
Moreover, as a consequence of \eqref{2.1} we have that the following identities hold
\begin{equation}\label{con1}
P_0^{r+1}\,Q_0^{s+1}=P_i^{r+1}\, Q_i^{s+1}=P_j^{r+1}\, Q_j^{s+1}=
P_0^{r+1}\, Q_j^{s+1},\quad \forall \;i\in I,\,j\in J,
\end{equation}
for the values of the functions $P_{\mu}$ and $Q_{\mu}$ at a solution $(z_{\alpha},m_{\alpha})$. In particular these identities lead to the
following expressions for the  constraints arising from the invariance of \eqref{3.1} under the action of $\bV_{rs}$.
\begin{teh}
If  $(z_{\alpha},m_{\alpha})$ is a solution of the string equations \eqref{3.1} then it satisfies the identities
\begin{equation}\label{con11}\everymath{\displaystyle}
\sum_{\alpha \in \{0\}\cup I}\oint_{\Gamma_{\alpha}}
\Big(\frac{z_{\alpha}}{n_{\alpha}}\Big)^s z_{\alpha}^{(r-s)n_{\alpha}}m_{\alpha}^{s+1}\d z_{\alpha}
+(-1)^r\frac{s+1}{r+1}n_0^{r-s}
\sum_{j \in J}\oint_{\Gamma_j}
\Big(\frac{z_j}{n_j}\Big)^r z_j^{(s-r)n_j}m_j^{r+1}\d z_j=0,
\end{equation}
for all $r,s\geq 0$.
\end{teh}
\begin{proof}
 From \eqref{3.4a}-\eqref{3.4ab} we find that \eqref{con1} takes the form
\begin{align}\label{con2}\everymath{\displaystyle}
\nonumber
z_0^{n_0(r+1)}\Big(\frac{1}{n_0}\frac{m_0}{z_0^{n_0-1}}\Big)^{s+1}
&=z_i^{n_i(r+1)}\Big(\frac{1}{n_i}\frac{m_i}{z_i^{n_i-1}}\Big)^{s+1}
=\Big(\frac{z_j^{n_j}}{n_0}\Big)^{s+1}\Big(-\frac{n_0}{n_j}\frac{m_j}{z_j^{n_j-1}}\Big)^{r+1}
\\\\
\nonumber
&=\frac{1}{n_0^{s+1}}E_1^{r+1}E_2^{s+1},\quad \forall \; i\in I,\,j\in J,
\end{align}
and we have that
\begin{align}\label{con3}\everymath{\displaystyle}
\nonumber
&z_0^{n_0(r+1)}\Big(\frac{1}{n_0}\frac{m_0}{z_0^{n_0-1}}\Big)^{s+1}\partial_p\log E_1
=z_i^{n_i(r+1)}\Big(\frac{1}{n_i}\frac{m_i}{z_i^{n_i-1}}\Big)^{s+1}\partial_p\log E_1
=\frac{(\partial_p E_1^{r+1})E_2^{s+1}}{(r+1)n_0^{s+1}},
\\\\
\nonumber
&\Big(\frac{z_j^{n_j}}{n_0}\Big)^{s+1}\Big(-\frac{n_0}{n_j}\frac{m_j}{z_j^{n_j-1}}\Big)^{r+1}\partial_p\log E_2
=-\frac{r+1}{s+1}z_0^{n_0(r+1)}\Big(\frac{1}{n_0}\frac{m_0}{z_0^{n_0-1}}\Big)^{s+1}\partial_p\log E_1+\partial_p\Big(\frac{E_1^{r+1}E_2^{s+1}}{(s+1)n_0^{s+1}} \Big)
,
\end{align}
for all $r,s\geq 0$ and $i\in I,\;j\in J$. Hence if we proceed as in the proof of \eqref{3.14a} and take into account that
\begin{align*}
\partial_p\log E_1&=n_0\partial_p\log z_0=n_i\partial_p\log z_i,\quad
\forall i\in I,\\
\partial_p\log E_2&=n_j\partial_p\log z_j,\quad
\forall j\in J,
\end{align*}
it is straightforward to prove that
\begin{align}\label{con4}\everymath{\displaystyle}
\nonumber
&\sum_{\alpha \in \{0\}\cup I}\oint_{\Gamma_{\alpha}}
z_{\alpha}^{n_{\alpha}(r+1)}\Big(\frac{1}{n_{\alpha}}
\frac{m_{\alpha}}{z_{\alpha}^{n_{\alpha}-1}}\Big)^{s+1}
n_{\alpha}\frac{\d
z_{\alpha}}{z_{\alpha}}-\sum_{j \in J}\frac{s+1}{r+1}\oint_{\Gamma_j}
\Big(\frac{z_j^{n_j}}{n_0}\Big)^{s+1}\Big(-\frac{n_0}{n_j}\frac{m_j}{z_j^{n_j-1}}\Big)^{r+1}
n_j\frac{\d
z_j}{z_j}\\\\
\nonumber
=& \oint_{\gamma}z_0^{n_0(r+1)}\Big(\frac{1}{n_0}\frac{m_0}{z_0^{n_0-1}}\Big)^{s+1}\partial_p\log E_1\d p
=\oint_{\gamma}\frac{(\partial_p E_1^{r+1})E_2^{s+1}}{(r+1)n_0^{s+1}}
\d p=0,
\end{align}
where $ \gamma:=\sum_{\alpha=0}^M \gamma_{\alpha}$. This proves that
the identities  \eqref{con11} hold.
\end{proof}

By evaluating the integrals of the left-hand side we obtain the symmetry constraints in terms of differential equations for the free-energy function $F=\log\tau$.

\subsubsection*{Examples}

For $r=s=0$ the equation \eqref{con4} reduces to \eqref{3.14a}, so that it implies
\[
\sum_{\alpha}t_{\alpha 0}=0.
\]
The cases $(r,s)=(1,0)$ and $(r,s)=(2,0)$ correspond to the Virasoro constraints induced by
$\bV_1$ and $\bV_2$, respectively, and  lead to the identities
\begin{align*}\everymath{\displaystyle}
&\sum_{\alpha \in \{0\}\cup I} \partial_{\alpha,n_{\alpha}}F-
\sum_{j \in J}\frac{n_0}{n_j}\Big(\sum_{n-n'=n_j-1}nt_{jn}
 \partial_{j,n'-1}F+n_jt_{jn_j}t_{j0}+\frac{1}{2}
 \sum_{n+n'=n_j }nn't_{jn} t_{jn'} \Big)=0,\\\\
&\sum_{\alpha \in \{0\}\cup I} \partial_{\alpha,2n_{\alpha}}F+
\sum_{j \in
J}\Big(\frac{n_0}{n_j}\Big)^2\Big(\sum_{n-n'-n''=2n_j-2}nt_{jn}
 (\partial_{j,n'-1}F)(\partial_{j,n''-1}F)+2\sum_{n-n'=2n_j-1}nt_{jn}t_{j0}
 \partial_{j,n'-1}F
 \\\\
 &+\sum_{n+n'-n''=2n_j-1}nn't_{jn}t_{jn'}
 \partial_{j,n''-1}F
 +2n_jt_{j\;2n_j}t_{j0}^2 + \sum_{n+n'=2n_j }nn't_{jn}
t_{jn'}t_{j0}
\\\\
&+ \frac{1}{3}
 \sum_{n+n'+n''=2n_j }nn'n''t_{jn} t_{jn'}t_{jn''}
  \Big)=0.
\end{align*}


\begin{thebibliography}{99}

\bibitem{h1}  B. A. Kuperschmidt and Yu. I.  Manin, Funk. Anal. Appl. I \textbf{11}, (3)
,31 (1977); II \textbf{17}, (1), 25 (1978).

\bibitem{h2}  V. E. Zakharov, Func. Anal. Priloz. \textbf{14}, 89-98 (1980);
Physica \textbf{3D}, 193-202 (1981).

\bibitem{h3}  \textit{Singular limits of dispersive waves }(eds. N. M.
Ercolani et al), Nato Adv. Sci. Inst. Ser. B Phys.
\textbf{320}, Plenum, New York (1994).

\bibitem{krich1} I. M. Krichever, Func. Anal. Appl. {\bf22} (1989) 200

\bibitem{krich2} I. M. Krichever, Commun. Pure. Appl. Math. {\bf 47} (1992) 437

\bibitem{kod}  S. Aoyama and Y. Kodama, Commun. Math. Phys. \textbf{182},
 (1996) 185

\bibitem{gib}  Y. Gibbons and S.P. Tsarev, Phys. Lett \textbf{258A},
(1999) 263

\bibitem{kon1} B. Konopelchenko , L. Martinez Alonso and O. Ragnisco,
J. Phys. A: Math. Gen.
{\bf 34}, (2001) 10209

\bibitem{kon2} B. Konopelchenko and L. Martinez Alonso, Phys. Lett. {\bf 286A},
(2001) 161

\bibitem{zab1} P. W. Wiegmann and P. B. Zabrodin,  Comm. Math.
Phys. {\bf 213} (2000) 523

\bibitem{zab2} M. Mineev-Weinstein, P. Wiegmann and A. Zabrodin,
Phys. Rev. Lett. {\bf 84} (2000) 5106

\bibitem{zab3} A. Boyarsky, A. Marsahakov, O. Ruchaysky,
P. Wiegmann and A. Zabrodin, Phys. Lett.B {\bf 515} (2001)
483

\bibitem{zab4} O. Agam, E. Bettelheim, P. Wiegmann and A. Zabrodin,
Phys. Rev. Lett. {\bf 88} (2002) 236801

\bibitem{zab5} I. Krichever, M. Mineev-Weinstein, P. Wiegmann and A. Zabrodin,
Physica D {\bf 198} (2004) 1

\bibitem{zab6} A. Zabrodin,
Teor. Mat. Fiz.  {\bf 142} (2005) 197


\bibitem{zab7} R. Teodorescu, E. Bettelheim, O. Agam, A. Zabrodin
and P. Wiegmann, Nuc. Phys. B {\bf 700} (2004) 521;
Nuc. Phys. B {\bf 704} (2005) 407


\bibitem{kaz} V. Kazakov and A. Marsahakov,
J. Phys. A {\bf 36} (2003) 3107,

\bibitem{tak1} K. Takasaki,  Commun. Math. Phys. {\bf 170} (1995) 101

\bibitem{tt1}  K. Takasaki and T. Takebe,  Lett. Math.
Phys. {\bf 23} (1991) 205
\bibitem{tt2} K. Takasaki and T. Takebe, Int. J. Mod. Phys. {\bf A7} Suppl.1 (1992) 889
\bibitem{tt3}  K. Takasaki and T. Takebe,  Lett. Math.
Phys. {\bf 28} (1993) 165
\bibitem{tt4}  K. Takasaki and T. Takebe,  Rev. Math.
Phys. {\bf 7} (1995) 743
\bibitem{mel1}  L. Martinez Alonso and M. Ma\~{n}as,  J. Math. Phys. {\bf 44} (2003)
3294
\bibitem{mel2} L. Martinez Alonso and E. Medina, J. Phys. A: Math. Gen. {\bf 37} (2004)
12005
\bibitem{mel3} L. Martinez Alonso and E. Medina, Phys. Lett. B {\bf 610}(2005)
227

\bibitem{mano}  M. Ma\~{n}as, E. Medina and L. Martinez Alonso, J. Phys. A: Math. Gen. {\bf 39} (2006)
2349

\bibitem{hen} P. Henrici, \emph{Applied and Computational Complex Analysis} {\bf Vol. 3} John Wiley and Sons, Inc. (1993)


\end{thebibliography}
\end{document}